\documentclass[twocolumn,astrosymb]{aastex7}

\newcommand{\WASP}{WASP-39\,b\,}

\newcommand{\HD}{HD\,209458\,b\,}
\newcommand{\KELT}{KELT-11\,b\,}

\usepackage{xcolor}
\definecolor{Lavender}{RGB}{181,126,220}

\usepackage{graphicx}	
\usepackage{float}
\usepackage{amsmath}	
\usepackage{amssymb}	
\usepackage{booktabs}
\usepackage{soul,xcolor}  
\usepackage{amssymb}
\usepackage{upgreek}
\usepackage{bbm}
\usepackage{dsfont}
\usepackage{sansmath}
\usepackage{mathrsfs}
\usepackage{yfonts}
\usepackage{euscript}

\def\d{{\rm d}}
\def\D{{\rm D}}

\def\grad-s{\mbox{\boldmath{$\nabla$}}_{\!\! s}\,}

\def\bff{{\bf f}}
\def\bfOmega{\mbox{\boldmath{$\Omega$}}}

\def\LD{L_{\scriptscriptstyle{D}}}

\def\DelTDN{\Delta T_{\!\scriptscriptstyle{{\rm DN}}}}

\def\gsim{\;\rlap{\lower 2.5pt\hbox{$\sim$}}\raise 1.5pt\hbox{$>$}\;}
\def\lsim{\;\rlap{\lower 2.5pt\hbox{$\sim$}}\raise 1.5pt\hbox{$<$}\;}

\def\grad{\mbox{\boldmath{$\nabla$}}}
\def\bfOmega{\mbox{\boldmath{$\Omega$}}}

\def\bfzeta{\mbox{\boldmath{$\zeta$}}}
\def\beq{\begin{equation}}
\def\eeq{\end{equation}}

\newcommand{\LR}{$\! L_{\scriptscriptstyle{R}}\,$}

\usepackage{color}
\definecolor{com_red}{rgb}{.8,.2,0.1}
\definecolor{com_turq}{rgb}{.0,.4,0.6}
\definecolor{com_burg}{rgb}{.45,.05,0.15}
\definecolor{ins_green}{rgb}{.1,.5,0.1}

\shorttitle{}
\shortauthors{Skinner \& Wei}

\graphicspath{./}

\begin{document}

\setstcolor{red}

\title{Bridging the Atmospheric Circulations of Hot and Warm Giant Exoplanets}

\author[0000-0002-5263-385X]{J. W. Skinner $^{\dagger,} $}
\affiliation{Division of Geological and Planetary Sciences, 
California Institute of Technology, 1200 E California Blvd, 
Pasadena, CA 91125, USA}
\affiliation{Martin A. Fisher School of Physics, Brandeis 
University, 415 South Street, Waltham, MA 02453, USA}
\email{$\dagger \,$ jskinner@caltech.edu}

\author[0009-0001-3400-6940]{S. Wei}
\affiliation{Martin A. Fisher School of Physics, Brandeis 
University, 415 South Street, Waltham, MA 02453, USA}
\email{swei@brandeis.edu}

\begin{abstract}
We perform high-resolution atmospheric flow simulations of hot and warm giant exoplanets that are tidally locked. 
The modeled atmospheres are representative of those on \KELT and \WASP\!, which possess markedly different equilibrium temperatures but reside in a similar dynamical regime: in this regime, their key dynamical numbers (e.g., Rossby and Froude numbers) are comparable.
Despite their temperature difference, both planets exhibit qualitatively similar atmospheric circulation patterns, which are characterized by turbulent equatorial flows, anticyclonic polar vortices, and large-scale Rossby waves that gives rise to quasi-zonal flows in the extra-tropics (i.e., near $\pm$\,$20^\circ{}$). 
Quantitative differences between \KELT and \WASP atmospheres reflect their different Rossby deformation scales, which influence the horizontal length scale of wave–-vortex interactions and the overall structure of the circulation.
\end{abstract}


\section{Introduction}

In atmospheric dynamics it is well understood that large-scale circulation is shaped by a combination of key parameters, rather than by a single factor such as temperature. 
This is important for giant exoplanets because they can have a wide range of physical and thermal forcing parameters, and therefore could exhibit many different atmospheric circulation patterns.
Most past atmospheric circulation studies of exoplanets have focused on the close-in, tidally synchronized ``hot'' (and ``ultra-hot'') Jupiters, because of their relative abundance and more observations \citep[e.g.,][]{Showman&Guillot_2002, Choetal_2003, Choetal_2008, Dobbs-Dixon&Lin_2008, Rauscher&Menou_2010, Thrastarson&Cho_2010, Heng&Showman_2014, Polichtchoukal_2014, Komacek&Showman_2016, Maynetal_2017, Mendetal18, Hammondetal_2021}. 
Hot Jupiters have high equilibrium temperatures of $T_{\rm eq}\sim 1600{-}2500 \, {\rm K}$, strong day--night temperature differences of $T_{\rm DN}\sim 500{-}1000 \, {\rm K}$, and short thermal relaxation timescales $\tau_r \sim {\rm hours}{-}{\rm days}$. 
These conditions lead to highly dynamic and complex atmospheres, featuring strong (typically azonal) equatorial flows and large vortices that lead to potentially observable signatures \citep[e.g.,][]{Choetal_2003, Thrastarson&Cho_2011, Choetal_2021, Skinneretal_2023, Changeatetal_2024, Kafleetal_2025}.

While the atmospheric dynamics of the solar system's cold giant planets and extrasolar systems' hot giant planets have been better studied, the intermediate class of more temperate giant planets remains essentially unexplored \citep[e.g.,][]{Cho_2008, medvedevetal_2013, Kieferetal_2024}.
These planets include those with $T_{\rm eq}\sim1000{-}1300 \, {\rm K}$, $T_{\rm DN}\lesssim 100 \, {\rm K}$, and thermal relaxation timescales $\mathcal{O}({\rm \, days})$ \citep{Faedietal_2011, Lineetal_2013, Wittenmyeretal_2022, Samraetal_2023, Damassoetal_2024}. 
Key dynamical parameters---including the Rossby number $R_o$ (ratio of inertial to Coriolis forces), Froude number $F_r$ (ratio of inertial to gravitational forces), and Rossby deformation scale $\LD$ (characteristic length scale traversed by a gravity wave in an inertial period) \citep{Holton_2004}---can lead to qualitatively different circulation patterns among different classes or types of giant planets \citep{Cho&Polvani_1996b, Choetal_2008}.

In this study, we examine the atmospheric dynamics of two giant exoplanets: a hot exoplanet, exemplified by \KELT \citep{Pepper_2017_kelt}, and a more temperate exoplanet, exemplified by \WASP \citep{Faedietal_2011}.
We choose these planets because, despite having different equilibrium temperatures, they are characterized by similar dynamical parameter values, such as the Rossby and Froude numbers and the Rossby deformation scale.
Our simulations demonstrate that, despite their different equilibrium temperatures, \KELT\!- and \WASP\!-like planets exhibit very similar large-scale circulation patterns. 
Both planets develop strong eastward (prograde) equatorial flows, large-scale Rossby waves, and persistent polar vortices.
The most notable difference arises from the difference in their Rossby deformation scales \citep{Gill_1982}.
These features are significant because they shape the global temperature distribution and may be observable through phase curves, emission maps, and transit spectroscopy \citep[e.g.,][]{Choetal_2003, Choetal_2021, Skinneretal_2023, Kafleetal_2025}.
Note that \KELT and \WASP have already been observed with Hubble and JWST, and they are key targets for the upcoming Ariel mission \citep{Tsiarasetal_2018, Changeatetal_2020, Edwards&Tinetti_2022, Feinsteinetal_2023, Edwardsetal_2024, Maetal_2025}.

\section{Methodology}

\subsection{Numerical Model}

The governing equations and numerical model used in this work are same as those in \citet{Skinner&Cho_2025}.
We therefore provide only a brief summary here and refer the reader to that work, and references therein, for more details. 
As in that work, we solve the primitive equations in relative vorticity--divergence--potential-temperature ($\zeta$--$\delta$--$\Theta$) form with pressure $p$ as the vertical coordinate.
The equations are solved using the highly accurate and well-tested pseudospectral code BoB \citep{Rivietal_2002, Scotetal_2004, Polichtchoukal_2014, Skinner&Cho_2020}
One of BoB's advantages is that it converges exponentially fast---i.e., its numerical error decreases exponentially with increasing resolution, rather than algebraically \citep[e.g.,][]{Boyd_2000}. 
We refer the reader to \cite{Skinner&Cho_2020}, \cite{Polichtchoukal_2014},  and \cite{Choetal_2015} for extensive convergence tests and inter-model comparisons.

The resolution of all the simulations presented in this work is identical---i.e., T341L50, which means the horizontal direction is represented by 341 total and 341 zonal wavenumbers in the Legendre expansion and the vertical coordinate is represented by $50$ layers spaced linearly over the range, $p \in [0.0, 5.0]$~bars; here 1 bar $= 10^5$~Pa. 
Note that our simulations satisfy the minimum horizontal resolution requirement for numerical convergence for this range:
we have performed a series of simulations to verify that the current resolution sufficiently represents the atmospheric circulations of the planets studied.
The simulations may not be converged for a much larger $p$-range, particularly if the layer number density is significantly higher than that here---e.g., 10$^2$ per MPa \citep{Skinner&Cho_2021}.
The prognostic variables, $\{\zeta, \delta, \Theta\}$, are defined on linearly spaced ``half-levels'' in $p \in [0.05, 4.95]$~bars. 
Note also that, while the exact $p$-range of the stratified atmosphere is not well known for giant exoplanets, our $p$-range is chosen to encompass the depth of the mid-infrared radiative transfer contribution function for both planets \citep{Changeatetal_2024}. 

For the time integration, we use a second-order leapfrog scheme with a stepsize of $\Delta t = (1/6000)\, \tau_p$, where $\tau_p$ is each planet's orbital period.
With this stepsize the Courant–Friedrichs–Lewy (CFL) number stays well below unity, typically under $0.1$ \citep{Courant_1967}.
To damp the computational mode introduced by the leapfrog integration, we use a Robert–Asselin time filter \citep{Robert_1966, Asselin_1972} with a filter coefficient of $\epsilon = 0.02$ \citep{Thrastarson&Cho_2011}.
We also apply a small $16^{\rm th}$-order hyperviscosity for $\{\zeta, \delta, \theta\}$ with viscosity coefficient $\nu = 1.5 \times 10^{-43}$ in units of $R^{16}\tau_p^{-1}$ to damp energy near the grid scale \citep[see, e.g.,][]{Polichtchoukal_2014, Skinner&Cho_2021}.
All simulations start from rest and evolve to $t = 300 \, \tau_p$ under the prescribed thermal forcing, which
is significantly longer than the dominant dynamical timescales
(see Section~\ref{sec:2.2} and Figure~\ref{fig:tp}). 
No other parameterizations---such as radiative transfer, chemistry, or dissipative processes (e.g., gravity wave or ion drag)---are included in our simulations.

\subsection{Physical Setup}\label{sec:2.2}

\begin{figure}
    \centering
    \includegraphics[width=1.0\linewidth]{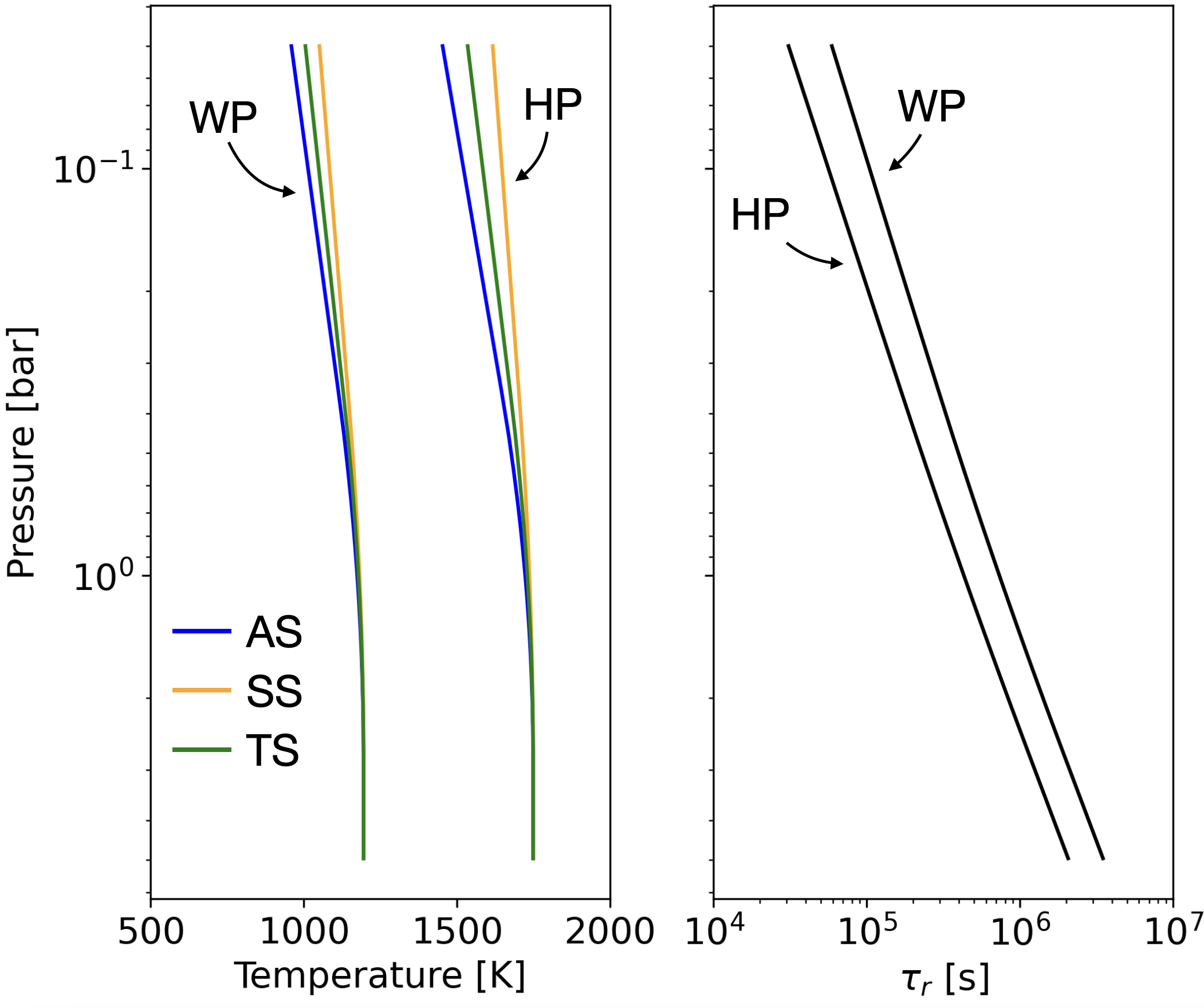}
    \caption{
    {\bf Left}: 
    Maximum and minimum equilibrium temperature profiles (yellow and blue lines, respectively) for the giant hot planet and warm planet (HP and WP, respectively). 
    Profiles at several different longitudes are shown: substellar point (SS), antistellar point (AS), and terminators (TS). 
    {\bf Right}: 
    Thermal relaxation timescale $\tau_{\rm r}(p)$ for HP and WP. 
    }
    \label{fig:tp}
\end{figure}

Some useful physical parameters and key dynamical numbers used for \KELT \citep{Changeatetal_2020} and \WASP \citep{Faedietal_2011}, our representative giant hot planet and warm planet (hereafter HP and WP, respectively), are listed in Table~\ref{tab:parameters}. 
Note that the HP has a $T_{\rm eq}$ similar to that of \HD, which is frequently used as a reference hot Jupiter; but, the former has a smaller $T_{\rm DN}$ in the thermally forced region of its modeled domain than the latter \citep[see Fig.~\ref{fig:tp} and][]{Changeatetal_2020}.  

\begin{table}
\centering
\setlength{\tabcolsep}{2pt}
\caption{
Useful physical parameters and dimensionless numbers for giant hot planet (HP) and warm planet (WP)---e.g., exoplanets 
\KELT and \WASP\!, respectively; 
the physical parameters for \KELT are from \citet{Pepper_2017_kelt} and for \WASP are from \citet{Faedietal_2011};
$^{*}$ at constant~$p$; 
$^{\dag}$ at $p = 0.1$~MPa; and,
$^\ddag$ characteristic values. 
}
\begin{tabular}{llll}
\toprule
\textbf{Parameters \hspace*{1cm}} & {\bf HP}  & {\bf WP} \hspace*{0.3cm}  & {\bf Units}\\
\midrule 
{\it Physical: } \\ 
$M$, planet mass & $3.7$ & 
$5.3$ & $\! 10^{26}$ Kg \\
$R$, \,\,planet radius & $9.8$ & 
$9.1$ & $\! 10^7$ m \\
$\Omega$,\ \  planet rot. rate & $1.5$ & 
$1.8$ & $\! 10^{-5}$ 1/s \\
$\tau_p$,\,\ planet rot. period & $4.1$ & 
$3.5$ & $\! 10^5$ s \\
$g$,\ \, surface gravity$^\dag$ & 2.5 & 4.1 & m/s$^2$ \\
$c_p$,\,\ specific heat$^{*}$, & $1.2$ & 
$1.2$ &  $\! 10^4$ m$^2$/(s$^2$\,K) \\
${\mathcal R}$,\,\ specific gas const. \hspace*{0.3cm} & 
$3.5$ \hspace*{0.3cm}  & 
$3.5$ \hspace*{0.3cm} & $\! 10^3$ m$^2$/(s$^2$\,K) \\
${\mathcal H}$,\, scale height$^\dag$ & 
$2.2$\, & $1.0$ & $\! 10^6$ m \\
$U$,\, max. wind speed$^\dag$  & $641$ & $529$ & 
m/s \\
${\mathcal N}$,\ Brunt--V\"ais\"al\"a freq. & 
$1.0$  & $2.0$ & $\! 10^{-3}$ 1/s \\
$c_s$,\,  adia. sound speed$^{\dag}$ & 
$2.9$\, & $2.4$\, & $\! 10^3$ m/s \\
\\
{\it Dynamical: } \\ 
$R_o$, Rossby number$^\ddag$  & 0.45 & 0.32 & -- \\
$F_r$,\ \,Froude number$^\ddag$  & $0.27$ & $0.26$ & -- \\
$B_u$,\ Burger number$^\ddag$  & $2.5$ & $1.5$ & -- \\
\LR,\ Deformation scale$^\ddag$ & $1.5$ & $1.1$ & $\, 10^8$ m\\ 
\bottomrule
\end{tabular}
\label{tab:parameters}
\end{table}

For the forcing setup, we use the commonly utilized Newtonian relaxation scheme \citep[e.g.,][]{Showman&Guillot_2002,Thrastarson&Cho_2010, Hengetal_2011, Liu&Showman_2013,Choetal_2015}.  
In this setup, the atmosphere is accelerated from a state at rest 
by a relaxation of the temperature field to specified dayside--nightside distribution (Figure~\ref{fig:tp} left, yellow and blue lines, respectively) 
over a prescribed relaxation time (Figure~\ref{fig:tp} right), as a function of $p$. 
The terminator profiles (Figure~\ref{fig:tp} left, green lines) serve as the initial condition for the simulations.
The Newtonian relaxation scheme is represented as a net heating rate in the PEs as 
$\dot{\mathcal{Q}} = -(T - T_{\rm eq})/ \tau_{\rm r}$, where 
$T = T(\lambda, \phi, p)$ is the local temperature at longitude $\lambda$ and latitude $\phi$, 
$T_{\rm eq} = T_{\rm eq}(\lambda, \phi, p)$ is
the equilibrium temperature, and 
$\tau_{\rm r} = \tau_{\rm r}(p)$ is the thermal relaxation time \citep[see, e.g.,][]{Choetal_2008}.
Laterally, $T_{\rm eq}(p)$ is modulated by $\cos\lambda\cos\phi$ on the planet's dayside as in \citet{Skinner&Cho_2021}.
Despite its idealized nature, the thermal forcing setup is useful for studying atmospheric dynamics when detailed physical properties are unknown.
As such, it remains a common approach in exoplanet modeling studies today \citep[e.g.,][]{Debras2020, boning2024, Skinner&Cho_2025, komacek_2025}.

To compute the $T_{\rm eq}(p)$ profiles in Figure~\ref{fig:tp}, we use a self-consistent approach for \KELT and \WASP which combines radiative transfer with an empirical relation derived from the population study of \cite{Changeatetal_2022}. 
A full description of our model, including refinements and validation steps, will be detailed elsewhere (Skinner et al., in prep.); see also \citet{Changeatetal_2024} for a brief summary/description.
Our model establishes a link between each planet’s atmospheric lapse rate and its retrieved dayside temperature using a second-order polynomial fit. 
The parameters of this fit are derived from statistical trends presented in \citet[][]{Changeatetal_2022}; see Figure 2 and Table 1 therein.
Since the exact $T_{\rm eq}(p)$ profiles for \KELT\ and \WASP\ are unknown, our method provides a physically plausible representation of their atmospheric thermal structures and ensures that both planets' profiles are derived in a consistent manner.
The thermal relaxation timescale $\tau_{\rm r}(p)$ is obtained from \citet{Cho_2008} for the profiles for the terminators of both planets. 

\section{Results}

\begin{figure*}
    \centering
    \includegraphics[width=0.98\linewidth]{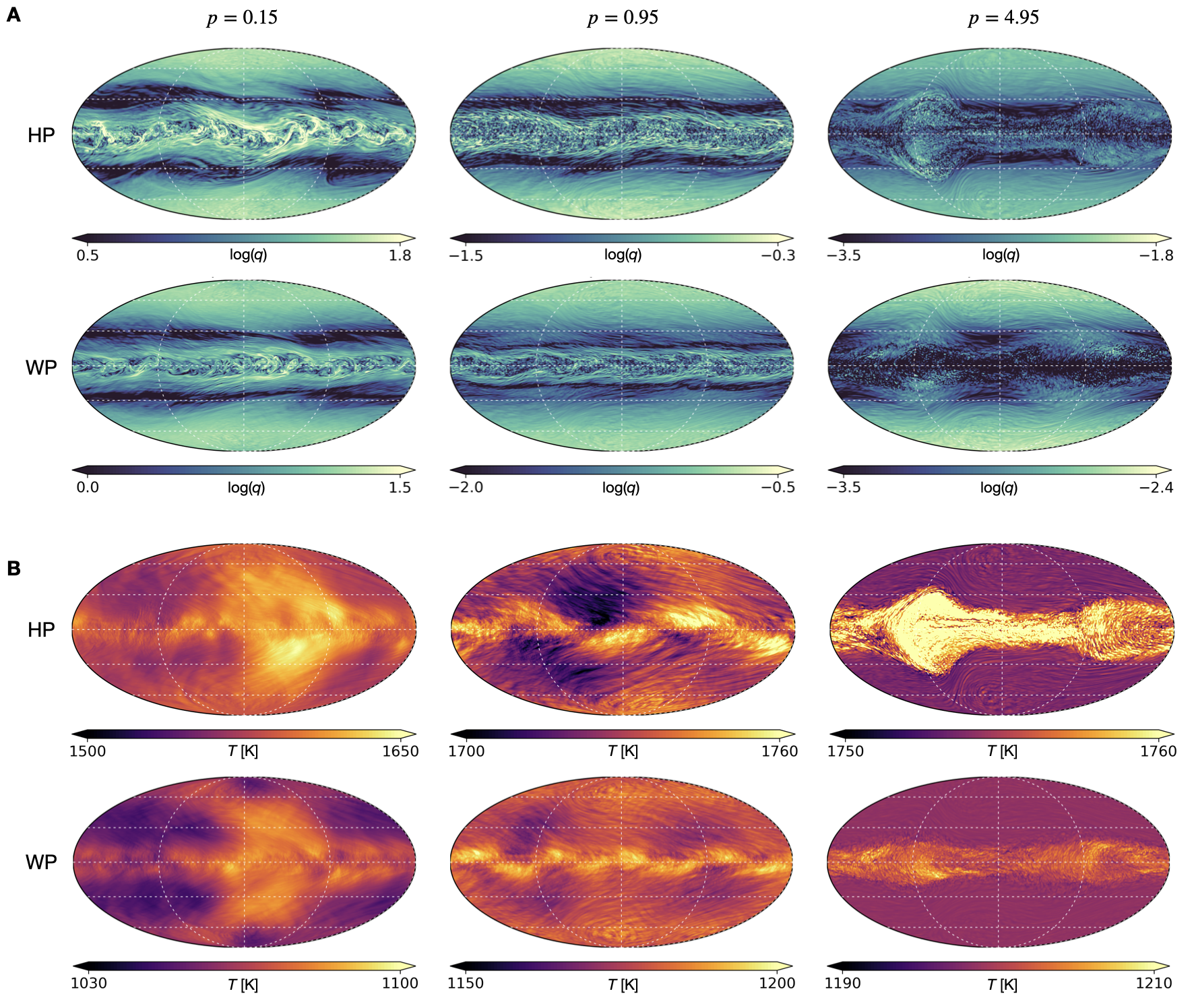}
    \caption{
    {\bf A}~$\log(|q(\lambda, \phi, p)|)$, where $q$ is the potential vorticity in units of $10^{-6}$ K m$^2$ kg$^{-1}$ s$^{-1}$, and {\bf B}~temperature $T(\lambda, \phi, p)$ fields in units of ${\rm K}$, at time $t = 200 \, \tau_p$ for T341L50 simulations of tidally synchronized hot and warm gas giant exoplanets (HP and WP, respectively).
    Fields are shown for $p \in \{0.15,\, 0.95,\, 4.95\}\,$bar.
    These $p$-levels are near the top, middle, and bottom of the simulation domains.
    Fields are shown in Mollweide projection, centered on the substellar point $(\lambda, \phi) = (0^\circ{}, 0^\circ{})$ where $\lambda$ is the longitude and $\phi$ is the latitude. 
    White horizontal dashed lines indicate the latitudes $\phi \in \{0^\circ{},\, \pm 30^\circ{},\, \pm 60^\circ{}\}$, and the white dashed circle indicates the dayside--nightside boundary.
    }
    \label{fig:zeta}
\end{figure*}

As can be seen in Figure~\ref{fig:tp} the thermal forcing is qualitatively the same for both planets. 
The forcing for both planets is characterized by a relatively modest\footnote{compared to hot-Jupiters such as \HD \citep[see, e.g.,][]{Skinner&Cho_2021}} dayside--nightside temperature difference 
$\DelTDN$ near the top of the modeled atmosphere and essentially no difference near the bottom of the modeled atmosphere: 
that is, $\DelTDN \sim 100$~K at $p = 0.05$~bar and decreases linearly in $\log (p)$ until $\DelTDN = 0$~K at $p = 1.0$~bar on both planets. 
Similarly, both planets have short relaxation timescales $\tau_{\rm r}$ near $p = 0.05$ which increase linearly in $\log(p)$ toward the bottom of the modeled domain.
Due to its lower mean $T_{\rm eq}$, \WASP (WP) has a larger $\tau_{\rm r}$ across the entire $p$ range compared to \KELT (HP).
Despite this quantitative difference in $T_{\rm eq}$ (hence in $\tau_{\rm r}$), their atmospheric circulation patterns are qualitatively similar---as we show below. 
This is because, despite the clear differences in physical or forcing parameters, the {\it{dynamical}} parameters of the two planets are very similar; see Table~\ref{tab:parameters}. 
This unique combination of shared dynamical parameters and quantitatively dissimilar physical properties provides a clear demonstration of the effect of dynamical parameters on atmospheric circulations.

Figure~\ref{fig:zeta} shows the atmospheric circulation patterns that emerge for both planets. 
The figure presents the potential vorticity \citep{Gill_1982} $q(\lambda, \phi, p)$, 

\begin{eqnarray}\label{eqn:pv}
    q\ =\ \frac{1}{\rho}\,(\bfzeta + \bff) \cdot \boldsymbol{\nabla}\Theta \, ,
\end{eqnarray}

and the temperature $T(\lambda, \phi, p)$ fields.
In equation~(\ref{eqn:pv}), $\rho$ is the density; $\nabla$ is the gradient operator;  $\bfzeta = \nabla \times \mathbf{v}$ is the relative vorticity; $\bff = 2\mathbf{\Omega}\sin \phi$ is the Coriolis parameter vector, where $\bfOmega$ is the planetary rotation vector that orients north; and, $\Theta$ is the potential temperature $\Theta = T \left(p_0/p \right)^{\kappa}$, where $p_0 = 5.0 \, {\rm bar}$ is the reference pressure and $\kappa = \mathcal{R}/c_p = 0.286$ is the ratio of the gas constant and specific heat at constant $p$. 
Under adiabatic conditions, $q$ is materially conserved (i.e., conserved following a fluid parcel) and serves as a tracer of the flow.
The $q$ fields for both planets are shown in $\log(|q|)$ scale, where $q$ is in potential vorticity units PVU (= $10^{-6}$ K m$^2$ kg$^{-1}$ s$^{-1}$).

Despite their markedly different mean $T_{\rm eq}$ (related to the irradiation received from the host star), both planets develop qualitatively similar atmospheric circulation patterns.
The circulations are governed by their modest $\Delta T_{\text{DN}}$ and short relaxation timescale in the upper region of the atmospheres ($p \lesssim 1$ bar) as well as the dynamical parameter values in the deeper, thermally unforced regions ($p \gtrsim 1$ bar).
By late simulation times ($t \gtrsim 100 \, \tau_p$), the atmospheres of both planets organize into a predominantly ``quasi-zonal'' circulation---particularly in the upper region of their atmospheres where the thermal forcing is applied (i.e., $p \lesssim 0.95\,$~bar);
here, by ``quasi-zonal'' we mean that the large-scale flow direction is predominantly in the east--west direction in the upper region of the atmosphere but is noticeably {\it a}zonal in some regions.
For example, near the equator north--south undulations as well as many small-scale coherent vortices can be seen. 
Moreover, in deeper regions $p \gtrsim 0.95\,$~bar and in the core of the equatorial flow, the meridional and zonal velocities have comparable amplitudes.
Note that these azonal structures are only captured in instantaneous snapshots of the fields and appear zonal when smoothed by averaging in space or time. 

\begin{figure}
    \centering
    \includegraphics[width=1\linewidth]{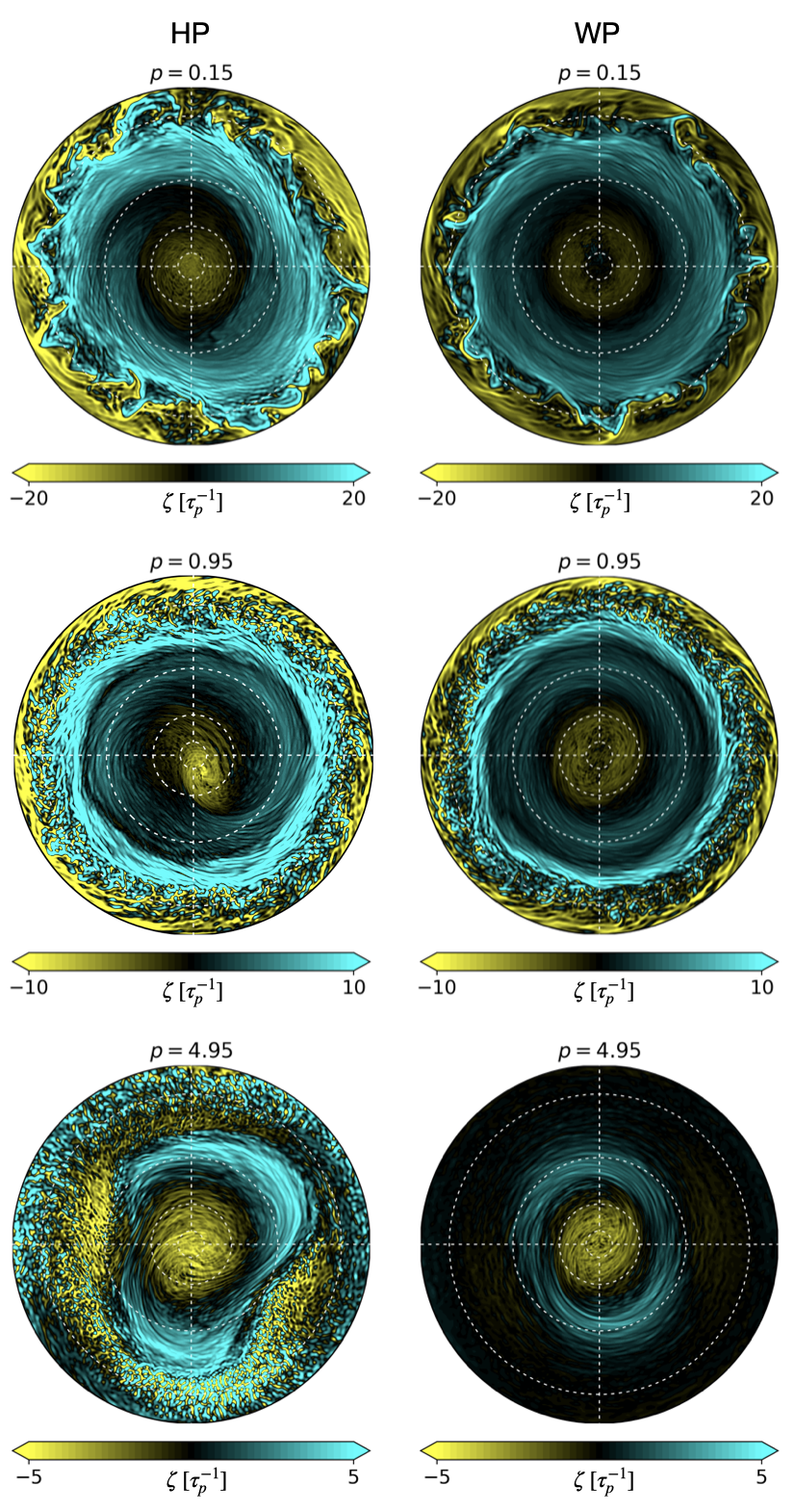}
     \caption{
     Relative vorticity $\zeta(\lambda, \phi, p)$ fields in units $\tau_p^{-1}$, in north polar stereographic projection at time $t = 170 \, \tau_p$.
     The plot boundaries are at $\phi = -10^\circ{}$ and radial dashed lines indicate the latitudes $\phi \in \{0^\circ{}, 30^\circ{}, 60^\circ{}, 80^\circ{}\}$.
     Both planets exhibit anticyclonic vortices at their poles surrounded by cyclonic flows near $\phi \sim 30^\circ{}$. 
     The HP's cyclonic flows are more distorted and elongated than those on the WP.  
     The WP's anticyclonic vortices are centered on the pole whereas the HP's are offset from the pole and precess around it (see $p = 0.95$~bar) with a period of $\sim 4 \tau_p$. 
     }
    \label{fig:pol}
\end{figure} 

It is also important to note that the emergence of the quasi-zonal flows in these simulations is due to the relatively weak thermal forcing applied. 
When we instead use the stronger and deeper-penetrating thermal forcing typical of \HD \citep[e.g.,][]{Skinner&Cho_2020}, while keeping the physical parameters of the HP and WP unchanged, the circulations become predominantly azonal. 
Under this forcing, the atmospheres also develop large-scale vortex pairs of opposite sign---consistent with the modon structures reported in, e.g., \citet{Skinner&Cho_2020}, \citet{Choetal_2021}, and \citet{Skinner&Cho_2021}.
Thus, the strength of the thermal forcing is important in setting the zonality of the atmospheric circulation.

Importantly, both planets also feature Rossby waves that distort their equatorial flows, particularly at their flanks.
Rossby waves are undulations of $q$-lines that result from the meridional gradient of $f(\phi)$.
These waves propagate westward relative to the mean flow.
On the HP, the undulations are more pronounced and the flanks of the equatorial flow extend further into the mid-latitudes. 
Planetary-scale patches of vorticity form between the equator and peripheries of the equatorial flow in the deeper regions of the atmosphere (see, e.g., $p = 4.95$ bar).

Key quantitative differences---such as the undulation magnitude just described---arise in a manner consistent with their different values of $\LD$.
The HP, with its larger $\LD$, develops broader equatorial flows that are $\sim$50\% stronger in magnitude than those on the WP.  
This can be seen in the peak values of $q$ in Fig.~\ref{fig:zeta}A.
Note that, in the figure, due to the logarithmic scale, the $0.2{-}0.3$ difference between the HP and WP means $q$ differs by a factor of $\sim$$1.6{-}2.0$.
Thus, the differences in $q$ (see, e.g., the equatorial flows at $p \in \{0.15, 0.95\}\,$ bar) represent a $60{–}100$\% difference in $q$ amplitude.
The WP, with its smaller $\LD$, develops narrower equatorial flows with sharper potential vorticity $q$ gradients along their flanks.
This is because $\LD$ sets the length scale over which flow structures interact in the atmosphere.
The smaller $\LD$ value associated with the WP concentrates $q$ gradients across smaller distances \citep{Dritschel&McIntyre_2008}.
Note that, while the mean $T_{\rm eq}$ does influence the $\LD$ (via the scale height $\mathcal{H}$ and the Brunt-V\"ais\"al\"a frequency $\mathcal{N}$), it is ultimately $\LD$ itself (via a combination of $\mathcal{H}$, $\mathcal{N}$, and $f$) that determines the outcome.

Figure~\ref{fig:zeta}B shows how the $T$ fields of both planets are correlated with their respective $q$ fields. 
This correlation arises because  $T$ is shaped by the same large-scale flow ($q$) structures---i.e., the quasi-zonal flows and vortices described above.
In regions where the Rossby number is small ($R_o \ll 1$), the flow is in quasi-geostrophic (QG) balance, a dynamical regime in which the Coriolis force is almost in balance with the horizontal pressure gradient.
In this regime, horizontal $T$ gradients are related to vertical shear in the zonal wind through thermal wind balance. 
The balance is met more strongly at the deeper pressure levels ($p \gtrsim 1$~bar).

The $T$ fields of both planets feature large (up to planetary length scale of $\sim$$R$) patches of hot and cold regions of the atmosphere.  
In general, hotter regions are located near to (but not at) the equator at all altitudes, while colder regions appear as large patches on the poleward flanks of the equatorial flow, around the mid-latitudes. 
{\it These thermal features are not stationary}, but drift westward over time, following the undulations of the equatorial flow.
The HP has a larger difference between the peak hot and cold regions at all $p$ levels, compared with the WP. 
The difference is $\sim$150~K at $p=0.15 \, {\rm bar}$ and $\sim$20~K at $p = 4.95 \, {\rm bar}$. 
The WP has a more moderate difference of $\lesssim 100 \, {\rm K}$ at $p=0.15 \, {\rm bar}$ and $\sim$10~K at $p = 4.95 \, {\rm bar}$.

\begin{figure*}
    \centering
    \includegraphics[width=0.95\linewidth]{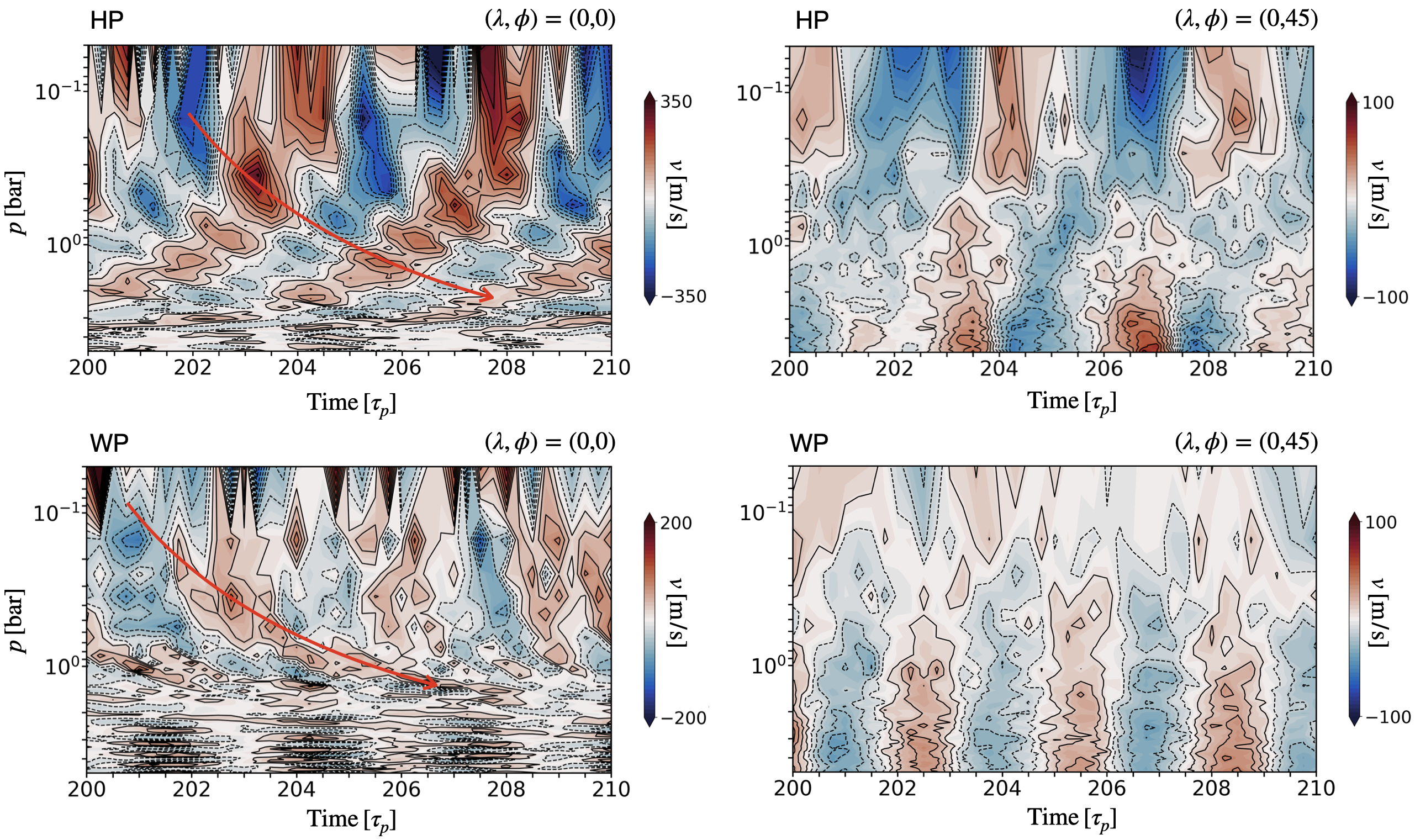}
    \caption{
        Hovm{\"o}ller plots showing the time evolution of meridional wind velocity $v$ as a function of pressure for the simulated hot planet (HP) and warm planet (WP) at the substellar point $(\lambda, \phi) = (0^\circ{},\,0^\circ{})$ (left) and mid-latitudes $(\lambda, \phi) = (0^\circ{}, \, 45^\circ{})$ (right) with red and blue colors corresponding to northward and southward flow, respectively. 
        Both planets exhibit oscillations in their equatorial regions with a period of $\lesssim$3\,$\tau_p$ and a downward propagation in time, as illustrated by red arrows.
        The HP has higher amplitude and more clear oscillations in $v$ compared to the WP. 
        Rossby waves appear as oscillations in the deep regions of both planets at $(\lambda, \phi) = (0^\circ{}, \, 45^\circ{})$.
        }
    \label{fig:hov}
\end{figure*}

In Fig.~\ref{fig:pol} we present a top-down look at both planets' $\zeta(\lambda, \phi, p)$ fields in the northern hemisphere. 
The fields are plotted in polar stereographic projection, centered on the north pole at $p \in \{0.15, 0.95, 4.95\}\,$~bar, at time $t = 170 \, \tau_p$, and between latitudes $\phi = -10^\circ{}$ (boundary of the plots) and $\phi = 90^\circ{}$ (center of the plots). 
Unlike in Fig.~\ref{fig:zeta}, these plots show cyclonic and anticyclonic regions as positive and negative $\zeta$ respectively; cyclonicity is defined by the sign of $\bfzeta \cdot \bfOmega$ (positive for cyclonic, negative for anticyclonic).

Both planets have {\it anticyclonic} vortices located at their poles, which are seen in Fig.~\ref{fig:pol} as regions of $\zeta < 0$ bounded by broad regions of $\zeta = 0$ at their periphery.
These polar vortices are encircled by {\it cyclonic} (as defined above) regions at the lower latitudes, identifiable in Fig.~\ref{fig:pol} as large patches of $\zeta > 0$ and also seen near the equator in Fig~\ref{fig:zeta}A. 
The bulk atmospheric motion of both planets is eastward near the equator and westward near the poles.
However, there are clear differences in the structure and evolution of the polar vortices and the equatorial flows between the two planets.
These differences are consistent with their different values of $\LD$ which controls how strongly the polar vortices interact with the equatorial flows \citep[e.g.,][]{Cho&Polvani_1996a}.

The HP has $\LD\!\sim\! 1.5R$, whereas $\LD\!\sim\! R$ for the WP.
Hence, the HP has stronger coupling between its equatorial and polar regions.
This results in the HP's polar vortices being displaced from the poles ($p=0.95\,$~bar, left column), with their center slowly orbiting the pole and completing one full orbit every $\sim$4$\, \tau_p$. 
This precession is clearly seen by visual inspection of movies of the planet's $\zeta$ fields.
In contrast, the WP's polar vortices remain centered on the poles and its cyclonic flows are generally more axisymmetric compared to those on the HP. 

In Fig.~\ref{fig:hov} we show Hovm{\"o}ller plots that illustrate the vertical structure and temporal evolution of the HP and WP atmospheres.
The plots show the meridional velocity $v(\lambda, \phi, p, t)$  (in units of m~s$^{-1}$) as a function of pressure $p$ and time (in units of $\tau_p$) at two locations: the substellar point $(\lambda, \phi) = (0^\circ{}, 0^\circ{})$ and at $(\lambda, \phi) = (0^\circ{}, 45^\circ{})$ of both planets.
The former location probes the equatorial flows and the latter probes just inside the periphery of the polar vortices shown in Fig.~\ref{fig:pol}.
Note the periphery of the polar vortex on the HP meanders and undulates much more than that on the WP.

Both planets exhibit oscillations with a period of $\sim$3\,$\tau_p$ with a transition in their behavior at the boundary of the thermally forced and unforced regions.
This transition occurs around $p \gtrsim 2$~bar for both planets.
In the thermally forced regions and at the substellar point for both planets, the oscillations appear to propagate downwards over time---i.e., from lower $p$ at early $t$ to higher $p$ later $t$, as illustrated by red curved arrows.
The oscillations are more clearly defined for the HP and have larger amplitude variations in $v$ (up to $\pm 350$\,m\,s$^{-1}$), compared with the WP's peak $v$ amplitude of $\pm 100$\,m\,s$^{-1}$.
In the thermally unforced regions at the substellar point, oscillations are aligned in $p$ for both planets.

Our simulations show that both planets develop prominent ``mode-2" Rossby waves in the deep, unforced regions of their atmospheres after approximately $t \sim 100 \, \tau_p$. 
By ``mode-2" we mean the Rossby waves have zonal wavenumber $m$ of 2. 
We estimate the wave phase speed $c_p$ of the Rossby waves from $c_p = \omega / m$, where $\omega$ is the observed angular frequency of the wave and $m = 2$.
Both are directly measured by detailed analyses of the fields.
A wave period of $\tau_\beta \sim 3 \,\tau_p$ is also obtained by visual inspection of Fig.~\ref{fig:hov}.

Both planets have $R_{\rm o} \lesssim 0.15$ and Burger number $B_{\rm u} = (\LD/L)^2 = \mathcal{O}(1)$ in their mid--to--high latitude regions (i.e., away from the equatorial flow) at essentially all $p$-levels.
Hence, we can apply the QG $\beta$-plane approximation to interpret the meridional structure of the Rossby waves in the thermally unforced region \citep{lipps_1964}. 
We calculate the dominant total wavenumber $n$ from the dispersion relation for baroclinic Rossby waves under the QG $\beta$-plane approximation:
\begin{equation}
    c_p\ =\ -\frac{\beta}{n(n+1)/R^2 + 1/\LD^2} \, .
    \label{eq:cp}
\end{equation}
Here, $\beta = R^{-1} df/d\phi = 2\Omega\cos\phi/R$ is the meridional gradient of $f$ on the sphere.
Using values of $\beta$, $c_p$ calculated from the physical parameters of both planets in Table~\ref{tab:parameters}, equation~\ref{eq:cp} predicts the dominant linear mode has $n \approx 3$.
With $m = 2$ and $n = 3$ the wave has four alternating longitudinal lobes (crests and troughs) in each hemisphere, organized into two latitudinal bands separated by a nodal line at the equator.
The presence of this structure is confirmed by visual inspection of the $v$ fields in the deeper atmospheric regions. 
Furthermore, the temporal propagation of this mode is seen as oscillations in the deep atmospheric regions for both planets in Fig.\ref{fig:hov} at $(\lambda, \phi) = (0^\circ, 45^\circ)$; see $p \gtrsim 2\,$~bar for the HP and $p\gtrsim 1$~bar for the WP.

\begin{figure}
    \centering
    \includegraphics[width=1.0\linewidth]{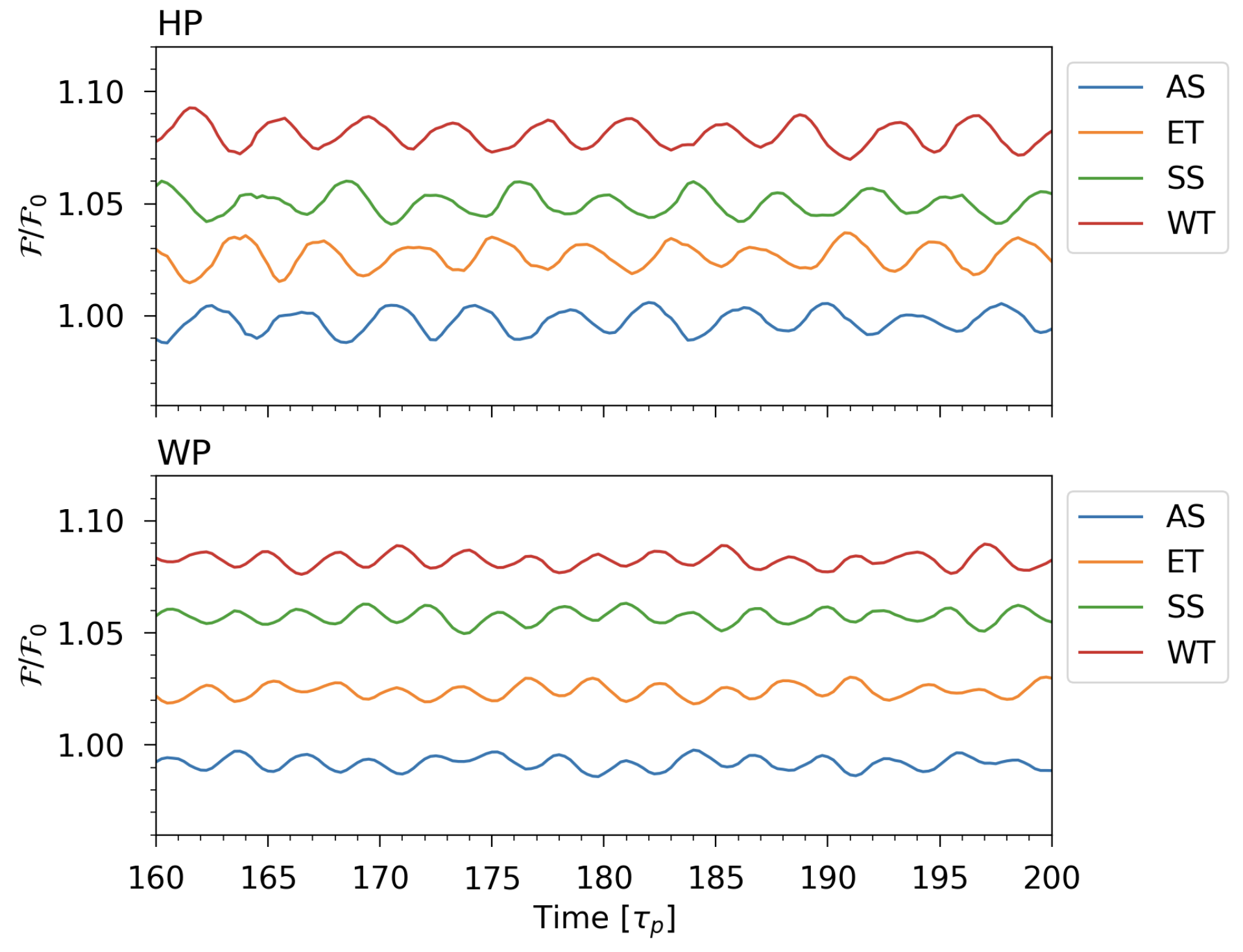}
    \caption{
    Normalized disc-averaged thermal flux $\mathcal{F}(t) / \mathcal{F}(t=0)$ at $p = 0.95\,$~bar, computed using the Stefan-Boltzmann law and weighted by a cosine function.
    The disc-average is centered on four key locations: the substellar point (SS), antistellar point (AS), eastern terminator (ET), and western terminator (WT).  
    These represent the emergent flux observed by JWST and Ariel \citep[see][]{Kafleetal_2025}.
    The lines are offset in the y-axis by $0.03$ and HP and WP are the hot and warm gas giant planets, respectively.  
    The HP has larger amplitude and shorter period oscillations than the WP. 
    }
    \label{fig:flux}
\end{figure}

In Figure~\ref{fig:flux} we illustrate the temporal behaviors of the HP and WP simulations for spatially averaged quantities relevant to observations.
The figure shows timeseries plots of normalized disk averaged thermal flux
\begin{equation}
\mathcal{F}(t) \equiv \sigma \langle T(\lambda, \phi, p, t)^4\rangle_\chi \cos \lambda \cos \phi
\end{equation}
at four key locations---the substellar point (SS), antistellar point (AS), eastern terminator (ET) and western terminator (WT)---and weighted by a cosine factor to account for the spherical geometry. 
The lines for the different locations are offset in the y-axis by $0.03$ for visual clarity.
Here, $\chi$ denotes the location of the disk center and $\langle \cdot \rangle$ the disk average.
For example, $\langle \cdot \rangle_{\rm SS}$ is the average centered on the substellar point. 
The Stefan-Boltzmann constant is $\sigma$, and emissivity is assumed to be zero.

To highlight relative temporal variations, the flux values $\mathcal{F}(t)$ are normalized by their initial values $\mathcal{F}(t=0)$ for each simulation. 
Since each simulation starts with a spatially uniform temperature distribution (Figure~\ref{fig:tp}, green lines), the normalization factor differs between the planets but remains independent of the disk-center location for each planet---i.e., the time series at all locations start with $\mathcal{F}/\mathcal{F}_0 = 1$ at $t = 0$.
A time window of $40 \, \tau_p$ is shown for $t \in [160, 200] \, \tau_p$, to illustrate the late time behavior of the simulations \citep[see e.g.,][]{Skinner&Cho_2025}.
The normalized flux $\mathcal{F}(t) / \mathcal{F}_0$ timeseries are shown at $p = 0.95 \,$~bar. 
The fluxes at this $p$ level match closely to full radiative transfer (RT) fluxes, so represent the emergent flux observed by JWST and Ariel \citep[see][]{Kafleetal_2025}.

Both planets have qualitatively the same normalized disk-averaged flux time series with periodic oscillations that are phase-shifted across locations and show no clear hotspot displacement (i.e., the different locations have comparable flux magnitudes that oscillate coherently).
The phase shifts indicate a westward propagation of hot and cold atmospheric regions on both planets, as can be seen by the progression of peaks and troughs in the time series across locations.
However, the flux timeseries differ quantitatively between the two planets.
The HP has higher amplitude oscillations with a slightly longer period of $\sim$4\,$\tau_p$ at all locations, consistent with the oscillations in Fig.~\ref{fig:hov} and precession timescale of the planet's polar vortex.
The WP has weaker, lower-amplitude fluctuations with period $\sim 3 \, \tau_p$ at all locations.

\section{Discussion}

In this paper, we have presented results from atmospheric dynamics simulations of tidally synchronized gas giants that are traditionally classified as ``hot'' and ``warm'' exoplanets. 
The hot exoplanet \KELT\,and the more temperate exoplanet \WASP\! exemplify the two classes.
These planets have markedly different mean equilibrium temperatures $T_{\rm eq}$ but have essentially the same dynamical parameters---except for one, the Rossby deformation scale $\LD$. 
Our simulations clearly demonstrate that, despite the significantly different values of $T_{\rm eq}$, both planets exhibit qualitatively the same large-scale circulation patterns; their circulations are characterized by quasi-zonal flows, large polar vortices, and high-amplitude Rossby waves. 
Our finding suggests that the qualitative features of large-scale atmospheric circulation on tidally-synchronized exoplanets is robust across a range of equilibrium temperatures, provided that their key dynamical parameters (such as the Rossby number, Froude number and Rossby deformation scale) remain comparable.

Quantitatively, the small difference in $\LD$ values between the two planets leads to distinct, differences in the circulations---indicating $\LD$ plays an important role the atmospheric response to thermal forcing on tidally-synchronized exoplanets.
On a planet that has the larger $\LD$ (e.g., \KELT, compared with \WASP), the atmospheric circulation is characterized by broader, more diffuse equatorial flows, precessing polar vortices, and stronger Rossby waves---as reported by \citet{Choetal_2003} and \citet{Choetal_2008}. 
This results in a more dynamically variable atmosphere with pronounced undulations and thermal variations that lead to larger amplitude signatures in disc-integrated flux.
In contrast, the circulation on \WASP, e.g.,---with the smaller $\LD$---is more zonal with lower-amplitude undulations.
Our finding underscores the important role of $\LD$ in influencing not only the spatial structure of the circulation, but also its temporal variability and the redistribution of hot and cold regions of atmosphere, which have direct implications for observable quantities such as phase curve amplitude and emitted flux.

\section{Acknowledgments} 
We thank Quentin Changeat and James Y-K. Cho for helpful discussions and suggestions.
This work used high performance computing at the San Diego Super Computing Center computing awarded to J.W.S through allocation PHY230189 from the Advanced Cyberinfrastructure Coordination Ecosystem: Services \& Support (ACCESS) program, which is supported by National Science Foundation grants \#2138259, \#2138286, \#2138307, \#2137603, and \#2138296 as well as high-performance computing awarded to J.W.S by the Google Cloud Research Credits program GCP19980904.

\bibliography{references}

\begin{thebibliography}{}
\expandafter\ifx\csname natexlab\endcsname\relax\def\natexlab#1{#1}\fi
\providecommand{\url}[1]{\href{#1}{#1}}
\providecommand{\dodoi}[1]{doi:~\href{http://doi.org/#1}{\nolinkurl{#1}}}
\providecommand{\doeprint}[1]{\href{http://ascl.net/#1}{\nolinkurl{http://ascl.net/#1}}}
\providecommand{\doarXiv}[1]{\href{https://arxiv.org/abs/#1}{\nolinkurl{https://arxiv.org/abs/#1}}}

\bibitem[{R. Asselin(1972)Asselin}]{Asselin_1972}
Asselin, R. 1972, \bibinfo{title}{{Frequency Filter for Time Integrations},} Monthly Weather Review, 100, 487, \dodoi{10.1175/1520-0493(1972)100<0487:FFFTI>2.3.CO;2}

\bibitem[{J.~P. Boyd(2000)Boyd}]{Boyd_2000}
Boyd, J.~P. 2000, {Chebyshev {\&} Fourier Spectral Methods}, 2nd edn. (Mineola, NY: Dover Publications)

\bibitem[{V.~G.~A. Böning {et~al.}(2024)Böning, Dietrich, \& Wicht}]{boning2024}
Böning, V. G.~A., Dietrich, W., \& Wicht, J. 2024, \bibinfo{title}{Westward hotspot offset explained by subcritical dynamo action in an ultra-hot Jupiter atmosphere,} \doarXiv{2407.12434}

\bibitem[{Q. Changeat {et~al.}(2020)Changeat, Edwards, Al-Refaie, Morvan, Tsiaras, Waldmann, \& Tinetti}]{Changeatetal_2020}
Changeat, Q., Edwards, B., Al-Refaie, A.~F., {et~al.} 2020, \bibinfo{title}{KELT-11 b: Abundances of water and constraints on carbon-bearing molecules from the hubble transmission spectrum,} The Astronomical Journal, 160, 260

\bibitem[{Q. Changeat {et~al.}(2022)Changeat, Edwards, Al-Refaie, Tsiaras, Skinner, Cho, Yip, Anisman, Ikoma, Bieger, Venot, Shibata, Waldmann, \& Tinetti}]{Changeatetal_2022}
Changeat, Q., Edwards, B., Al-Refaie, A.~F., {et~al.} 2022, \bibinfo{title}{{Five Key Exoplanet Questions Answered via the Analysis of 25 Hot-Jupiter Atmospheres in Eclipse},} The Astrophysical Journal Supplement Series, 260, 3, \dodoi{10.3847/1538-4365/ac5cc2}

\bibitem[{Q. Changeat {et~al.}(2024)Changeat, Skinner, Cho, Nättilä, Waldmann, Al-Refaie, Dyrek, Edwards, Mikal-Evans, Joshua, Morello, Skaf, Tsiaras, Venot, \& Yip}]{Changeatetal_2024}
Changeat, Q., Skinner, J.~W., Cho, J.~{\relax Y-K}., {et~al.} 2024, \bibinfo{title}{Is the Atmosphere of the Ultra-hot Jupiter WASP-121 b Variable?} The Astrophysical Journal Supplement Series, 270, 34, \dodoi{10.3847/1538-4365/ad1191}

\bibitem[{J.~Y. Cho \& L.~M. Polvani(1996)Cho \& Polvani}]{Cho&Polvani_1996b}
Cho, J.~Y., \& Polvani, L.~M. 1996, \bibinfo{title}{The morphogenesis of bands and zonal winds in the atmospheres on the giant outer planets,} Science, 273, 335

\bibitem[{J.~{\relax Y-K}. Cho(2008)Cho}]{Cho_2008}
Cho, J.~{\relax Y-K}. 2008, \bibinfo{title}{{Atmospheric dynamics of tidally synchronized extrasolar planets},} Philosophical Transactions of the Royal Society A: Mathematical, Physical and Engineering Sciences, 366, \dodoi{10.1098/rsta.2008.0177}

\bibitem[{J.~{\relax Y-K}. Cho {et~al.}(2003)Cho, Menou, Hansen, \& Seager}]{Choetal_2003}
Cho, J.~{\relax Y-K}., Menou, K., Hansen, B. M.~S., \& Seager, S. 2003, \bibinfo{title}{{Changing Face of the Extrasolar Giant Planet, HD 209458b},} The Astrophysical Journal Letters, 587, 117, \dodoi{10.1086/375016}

\bibitem[{J.~{\relax Y-K}. Cho {et~al.}(2008)Cho, Menou, Hansen, \& Seager}]{Choetal_2008}
Cho, J.~{\relax Y-K}., Menou, K., Hansen, B. M.~S., \& Seager, S. 2008, \bibinfo{title}{{Atmospheric Circulation of Close‐in Extrasolar Giant Planets. I. Global, Barotropic, Adiabatic Simulations},} The Astrophysical Journal, 675, 817, \dodoi{10.1086/524718}

\bibitem[{J.~{\relax Y-K}. Cho {et~al.}(2015)Cho, Polichtchouk, \& Thrastarson}]{Choetal_2015}
Cho, J.~{\relax Y-K}., Polichtchouk, I., \& Thrastarson, H.~{\relax Th}. 2015, \bibinfo{title}{{Sensitivity and variability redux in hot-Jupiter flow simulations},} Monthly Notices of the Royal Astronomical Society, 454, 3423, \dodoi{10.1093/mnras/stv1947}

\bibitem[{J.~{\relax Y-K}. Cho \& L.~M. Polvani(1996)Cho \& Polvani}]{Cho&Polvani_1996a}
Cho, J.~{\relax Y-K}., \& Polvani, L.~M. 1996, \bibinfo{title}{{The emergence of jets and vortices in freely evolving, shallow-water turbulence on a sphere},} Physics of Fluids, 8, 1531, \dodoi{10.1063/1.868929}

\bibitem[{J.~{\relax Y-K}. Cho {et~al.}(2021)Cho, Skinner, \& Thrastarson}]{Choetal_2021}
Cho, J.~{\relax Y-K}., Skinner, J.~W., \& Thrastarson, H.~{\relax Th}. 2021, \bibinfo{title}{{Storms, Variability, and Multiple Equilibria on Hot Jupiters},} The Astrophysical Journal Letters, 913, 832, \dodoi{10.3847/2041-8213/abfd37}

\bibitem[{R. Courant {et~al.}(1967)Courant, Friedrichs, \& Lewy}]{Courant_1967}
Courant, R., Friedrichs, K., \& Lewy, H. 1967, \bibinfo{title}{On the partial difference equations of mathematical physics,} IBM journal of Research and Development, 11, 215

\bibitem[{M. Damasso {et~al.}(2024)Damasso, Polychroni, Locci, Turrini, Maggio, Cubillos, Baratella, Biazzo, Benatti, Mantovan, {et~al.}}]{Damassoetal_2024}
Damasso, M., Polychroni, D., Locci, D., {et~al.} 2024, \bibinfo{title}{TOI-837 b: Characterisation, formation, and evolutionary history of an infant warm Saturn-mass planet,} Astronomy \& Astrophysics, 688, A15

\bibitem[{F. Debras {et~al.}(2020)Debras, Mayne, Baraffe, Jaupart, Mourier, Laibe, Goffrey, \& Thuburn}]{Debras2020}
Debras, F., Mayne, N., Baraffe, I., {et~al.} 2020, \bibinfo{title}{Acceleration of superrotation in simulated hot {Jupiter} atmospheres,} Astronomy \& Astrophysics, 633, A2, \dodoi{10.1051/0004-6361/201936110}

\bibitem[{I. Dobbs‐Dixon \& D.~N.~C. Lin(2008)Dobbs‐Dixon \& Lin}]{Dobbs-Dixon&Lin_2008}
Dobbs‐Dixon, I., \& Lin, D. N.~C. 2008, \bibinfo{title}{{Atmospheric Dynamics of Short‐Period Extrasolar Gas Giant Planets. I. Dependence of Nightside Temperature on Opacity},} The Astrophysical Journal, 673, \dodoi{10.1086/523786}

\bibitem[{D. Dritschel \& M. McIntyre(2008)Dritschel \& McIntyre}]{Dritschel&McIntyre_2008}
Dritschel, D., \& McIntyre, M. 2008, \bibinfo{title}{Multiple jets as PV staircases: the Phillips effect and the resilience of eddy-transport barriers,} Journal of the Atmospheric Sciences, 65, 855

\bibitem[{B. Edwards \& G. Tinetti(2022)Edwards \& Tinetti}]{Edwards&Tinetti_2022}
Edwards, B., \& Tinetti, G. 2022, \bibinfo{title}{The Ariel Target List: The Impact of TESS and the Potential for Characterizing Multiple Planets within a System,} The Astronomical Journal, 164, 15

\bibitem[{B. Edwards {et~al.}(2024)Edwards, Tsiaras, Changeat, \& Yip}]{Edwardsetal_2024}
Edwards, B., Tsiaras, A., Changeat, Q., \& Yip, K.~H. 2024, \bibinfo{title}{On the difficulties of obtaining absolute transit depths with HST WFC3: KELT-11 b, an example,} RAS Techniques and Instruments, 3, 415

\bibitem[{F. Faedi {et~al.}(2011)Faedi, Barros, Anderson, Brown, Cameron, Pollacco, Boisse, Hebrard, Lendl, Lister, {et~al.}}]{Faedietal_2011}
Faedi, F., Barros, S.~C., Anderson, D.~R., {et~al.} 2011, \bibinfo{title}{WASP-39b: a highly inflated Saturn-mass planet orbiting a late G-type star,} Astronomy \& Astrophysics, 531, A40

\bibitem[{A.~D. Feinstein {et~al.}(2023)Feinstein, Radica, Welbanks, Murray, Ohno, Coulombe, Espinoza, Bean, Teske, Benneke, {et~al.}}]{Feinsteinetal_2023}
Feinstein, A.~D., Radica, M., Welbanks, L., {et~al.} 2023, \bibinfo{title}{Early Release Science of the exoplanet WASP-39b with JWST NIRISS,} Nature, 614, 670

\bibitem[{A. Gill(1982)Gill}]{Gill_1982}
Gill, A. 1982, Atmosphere-Ocean Dynamics, Atmosphere-Ocean Dynamics No. v. 30 (Elsevier Science)

\bibitem[{M. Hammond \& N.~T. Lewis(2021)Hammond \& Lewis}]{Hammondetal_2021}
Hammond, M., \& Lewis, N.~T. 2021, \bibinfo{title}{The rotational and divergent components of atmospheric circulation on tidally locked planets,} Proceedings of the National Academy of Sciences, 118, e2022705118

\bibitem[{K. {Heng} {et~al.}(2011){Heng}, {Menou}, \& {Phillipps}}]{Hengetal_2011}
{Heng}, K., {Menou}, K., \& {Phillipps}, P.~J. 2011, \bibinfo{title}{{Atmospheric circulation of tidally locked exoplanets: a suite of benchmark tests for dynamical solvers},} \mnras, 413, 2380, \dodoi{10.1111/j.1365-2966.2011.18315.x}

\bibitem[{K. Heng \& A.~P. Showman(2014)Heng \& Showman}]{Heng&Showman_2014}
Heng, K., \& Showman, A.~P. 2014, \bibinfo{title}{{Atmospheric Dynamics of Hot Exoplanets},} \dodoi{10.1146/annurev-earth-060614-105146}

\bibitem[{J.-R. Holton(2004)Holton}]{Holton_2004}
Holton, J.-R. 2004, {An Introduction to Dynamic Meteorology}, 4th edn. (Elsevier Academic Press)

\bibitem[{J. Kafle {et~al.}(2025)Kafle, Cho, \& Changeat}]{Kafleetal_2025}
Kafle, J., Cho, J.~Y., \& Changeat, Q. 2025, \bibinfo{title}{Radiative Flux from a High-Resolution Atmospheric Dynamics Simulation of a Hot-Jupiter for JWST and Ariel,} arXiv preprint arXiv:2504.11679

\bibitem[{S. Kiefer {et~al.}(2024)Kiefer, Bach-M{\o}ller, Samra, Lewis, Schneider, Amadio, Lecoq-Molinos, Carone, Decin, J{\o}rgensen, {et~al.}}]{Kieferetal_2024}
Kiefer, S., Bach-M{\o}ller, N., Samra, D., {et~al.} 2024, \bibinfo{title}{Under the magnifying glass: A combined 3D model applied to cloudy warm Saturn-type exoplanets around M dwarfs,} Astronomy \& Astrophysics, 692, A222

\bibitem[{T.~D. Komacek(2025)Komacek}]{komacek_2025}
Komacek, T.~D. 2025, \bibinfo{title}{Limited hysteresis in the atmospheric dynamics of hot Jupiters,} arXiv preprint arXiv:2502.19394

\bibitem[{T.~D. Komacek \& A.~P. Showman(2016)Komacek \& Showman}]{Komacek&Showman_2016}
Komacek, T.~D., \& Showman, A.~P. 2016, \bibinfo{title}{Atmospheric circulation of hot Jupiters: dayside--nightside temperature differences,} The Astrophysical Journal, 821, 16

\bibitem[{M.~R. Line {et~al.}(2013)Line, Knutson, Deming, WILkINS, \& Desert}]{Lineetal_2013}
Line, M.~R., Knutson, H., Deming, D., WILkINS, A., \& Desert, J.-M. 2013, \bibinfo{title}{A near-infrared transmission spectrum for the warm Saturn HAT-P-12b,} The Astrophysical Journal, 778, 183

\bibitem[{F.~B. Lipps(1964)Lipps}]{lipps_1964}
Lipps, F.~B. 1964, \bibinfo{title}{A note on the beta-plane approximation,} Tellus, 16, 535

\bibitem[{B. Liu \& A.~P. Showman(2013)Liu \& Showman}]{Liu&Showman_2013}
Liu, B., \& Showman, A.~P. 2013, \bibinfo{title}{{Atmospheric circulation of hot jupiters: Insensitivity to initial conditions},} The Astrophysical Journal, 770, \dodoi{10.1088/0004-637X/770/1/42}

\bibitem[{S. Ma {et~al.}(2025)Ma, Saba, Al-Refaie, Tinetti, Yurchenko, Tennyson, \& Pestellini}]{Maetal_2025}
Ma, S., Saba, A., Al-Refaie, A.~F., {et~al.} 2025, \bibinfo{title}{A new look into the atmospheric composition of WASP-39 b,} arXiv preprint arXiv:2504.07823

\bibitem[{N.~J. Mayne {et~al.}(2017)Mayne, Debras, Baraffe, Thuburn, Amundsen, Acreman, Smith, Browning, Manners, \& Wood}]{Maynetal_2017}
Mayne, N.~J., Debras, F., Baraffe, I., {et~al.} 2017, \bibinfo{title}{{Astrophysics Results from a set of three-dimensional numerical experiments of a hot Jupiter atmosphere},} A{\&}A, 604, 79, \dodoi{10.1051/0004-6361/201730465}

\bibitem[{A.~S. Medvedev {et~al.}(2013)Medvedev, Sethunadh, \& Hartogh}]{medvedevetal_2013}
Medvedev, A.~S., Sethunadh, J., \& Hartogh, P. 2013, \bibinfo{title}{From cold to warm gas giants: A three-dimensional atmospheric general circulation modeling,} Icarus, 225, 228

\bibitem[{J.~M. Mendon{\c{c}}a {et~al.}(2018)Mendon{\c{c}}a, Tsai, Malik, Grimm, \& Heng}]{Mendetal18}
Mendon{\c{c}}a, J.~M., Tsai, S.-m., Malik, M., Grimm, S.~L., \& Heng, K. 2018, \bibinfo{title}{{Three-dimensional Circulation Driving Chemical Disequilibrium in WASP-43b},} The Astrophysical Journal, 869, 107, \dodoi{10.3847/1538-4357/aaed23}

\bibitem[{J. Pepper {et~al.}(2017)Pepper, Rodriguez, Collins, Johnson, Fulton, Howard, Beatty, Stassun, Isaacson, Col{\'o}n, {et~al.}}]{Pepper_2017_kelt}
Pepper, J., Rodriguez, J.~E., Collins, K.~A., {et~al.} 2017, \bibinfo{title}{{KELT-11B: a highly inflated sub-saturn exoplanet transiting the V= 8 Subgiant HD 93396},} The Astronomical Journal, 153, 215

\bibitem[{I. Polichtchouk {et~al.}(2014)Polichtchouk, Cho, Watkins, Thrastarson, Umurhan, \& de~la Torre-Ju{\'{a}}rez}]{Polichtchoukal_2014}
Polichtchouk, I., Cho, J.~{\relax Y-K}., Watkins, C., {et~al.} 2014, \bibinfo{title}{{Intercomparison of general circulation models for hot extrasolar planets},} Icarus, 229, 355, \dodoi{10.1016/j.icarus.2013.11.027}

\bibitem[{E. Rauscher \& K. Menou(2010)Rauscher \& Menou}]{Rauscher&Menou_2010}
Rauscher, E., \& Menou, K. 2010, \bibinfo{title}{{Three Dimensional Modeling of Hot Jupiter Atmospheric Flows},} The Astrophysical Journal, 714, 1334, \dodoi{10.1088/0004-637X/714/2/1334}

\bibitem[{L. Rivier {et~al.}(2002)Rivier, Loft, \& Polvani}]{Rivietal_2002}
Rivier, L., Loft, R., \& Polvani, L.~M. 2002, \bibinfo{title}{{An Efficient Spectral Dynamical Core for Distributed Memory Computers},} Monthly Weather Review, 130, \dodoi{10.1175/1520-0493(2002)130<1384:AESDCF>2.0.CO;2}

\bibitem[{A.~J. Robert(1966)Robert}]{Robert_1966}
Robert, A.~J. 1966, \bibinfo{title}{{The Integration of a Low Order Spectral Form of the Primitive Meteorological Equations},} Journal of the Meteorological Society of Japan. Ser. II, 44, 237, \dodoi{10.2151/jmsj1965.44.5_237}

\bibitem[{D. Samra {et~al.}(2023)Samra, Helling, Chubb, Min, Carone, \& Schneider}]{Samraetal_2023}
Samra, D., Helling, C., Chubb, K., {et~al.} 2023, \bibinfo{title}{Clouds form on the hot Saturn JWST ERO target WASP-96b,} Astronomy \& Astrophysics, 669, A142

\bibitem[{R.~K. Scott {et~al.}(2004)Scott, Rivier, Loft, \& Polvani}]{Scotetal_2004}
Scott, R.~K., Rivier, L., Loft, R., \& Polvani, L.~M. 2004, \bibinfo{title}{{BOB: Model Description and User Guide},} NCAR Technical Report, 456, \dodoi{doi:10.5065/D698850K}

\bibitem[{A.~P. Showman \& T. Guillot(2002)Showman \& Guillot}]{Showman&Guillot_2002}
Showman, A.~P., \& Guillot, T. 2002, \bibinfo{title}{{Atmospheric circulation and tides of “51 Pegasus b-like” planets},} Astronomy {\&} Astrophysics, 385, 166, \dodoi{10.1051/0004-6361:20020101}

\bibitem[{J.~W. Skinner \& J.~Y.-K. Cho(2025)Skinner \& Cho}]{Skinner&Cho_2025}
Skinner, J.~W., \& Cho, J. Y.-K. 2025, \bibinfo{title}{Early Time Small-scale Structures in Hot Exoplanet Atmosphere Simulations,} The Astrophysical Journal, 982, 64

\bibitem[{J.~W. Skinner \& J.~{\relax Y-K}. Cho(2021)Skinner \& Cho}]{Skinner&Cho_2020}
Skinner, J.~W., \& Cho, J.~{\relax Y-K}. 2021, \bibinfo{title}{{Numerical convergence of hot-Jupiter atmospheric flow solutions},} Monthly Notices of the Royal Astronomical Society, 504, 5172, \dodoi{10.1093/mnras/stab971}

\bibitem[{J.~W. Skinner \& J.~{\relax Y-K}. Cho(2022)Skinner \& Cho}]{Skinner&Cho_2021}
Skinner, J.~W., \& Cho, J.~{\relax Y-K}. 2022, \bibinfo{title}{{Modons on tidally synchronized extrasolar planets},} Monthly Notices of the Royal Astronomical Society, 511, 3584, \dodoi{10.1093/mnras/stab2809}

\bibitem[{J.~W. Skinner {et~al.}(2023)Skinner, N\"attil\"a, \& Cho}]{Skinneretal_2023}
Skinner, J.~W., N\"attil\"a, J., \& Cho, J.~{\relax Y-K}. 2023, \bibinfo{title}{Repeated Cyclogenesis on Hot-Exoplanet Atmospheres with Deep Heating,} Phys. Rev. Lett., 131, 231201, \dodoi{10.1103/PhysRevLett.131.231201}

\bibitem[{H.~{\relax Th}. Thrastarson \& J.~{\relax Y-K}. Cho(2010)Thrastarson \& Cho}]{Thrastarson&Cho_2010}
Thrastarson, H.~{\relax Th}., \& Cho, J.~{\relax Y-K}. 2010, \bibinfo{title}{{Effects of Initial Flow on Close-In Planet Atmospheric Circulation},} The Astrophysical Journal, 716, 144, \dodoi{10.1088/0004-637X/716/1/144}

\bibitem[{H.~{\relax Th}. Thrastarson \& J.~{\relax Y-K}. Cho(2011)Thrastarson \& Cho}]{Thrastarson&Cho_2011}
Thrastarson, H.~{\relax Th}., \& Cho, J.~{\relax Y-K}. 2011, \bibinfo{title}{{Relaxation time and dissipation interaction in hot planet atmospheric flow simulations},} The Astrophysical Journal, 729, 117, \dodoi{10.1088/0004-637X/729/2/117}

\bibitem[{A. Tsiaras {et~al.}(2018)Tsiaras, Waldmann, Zingales, Rocchetto, Morello, Damiano, Karpouzas, Tinetti, McKemmish, Tennyson, {et~al.}}]{Tsiarasetal_2018}
Tsiaras, A., Waldmann, I., Zingales, T., {et~al.} 2018, \bibinfo{title}{A population study of gaseous exoplanets,} The Astronomical Journal, 155, 156

\bibitem[{R.~A. Wittenmyer {et~al.}(2022)Wittenmyer, Clark, Trifonov, Addison, Wright, Stassun, Horner, Lowson, Kielkopf, Kane, {et~al.}}]{Wittenmyeretal_2022}
Wittenmyer, R.~A., Clark, J.~T., Trifonov, T., {et~al.} 2022, \bibinfo{title}{TOI-1842b: A Transiting Warm Saturn Undergoing Reinflation around an Evolving Subgiant,} The Astronomical Journal, 163, 82

\end{thebibliography}
\bibliographystyle{aasjournal}
\end{document}